\documentstyle[multicol,prl,aps,graphicx]{revtex}

\newcommand{\sqrts}{\mbox{$\sqrt{s}$}}
\newcommand{\sqrtsNN}{\mbox{$\sqrt{s_{_{\mathrm{NN}}}}$}}
\newcommand{\AuAu}{Au+Au}
\newcommand{\PbPb}{Pb+Pb}
\newcommand{\ppbar}{\mbox{$p\bar{p}$}}

\newcommand{\hminus}{\mbox{$h^-$}}
\newcommand{\Nhminus}{\mbox{$N_{h^-}$}}
\newcommand{\kzeros}{\mbox{$K^0_S$}}
\newcommand{\piminus}{\mbox{$\pi^-$}}
\newcommand{\kminus}{\mbox{$K^-$}}
\newcommand{\pbar}{\mbox{$\bar{p}$}}
\newcommand{\meanpt}{\mbox{$\langle p_\perp \rangle$}}
\newcommand{\pt}{\mbox{$p_\perp$}}
\newcommand{\meanEt}{\mbox{$\langle E_\perp \rangle$}}
\newcommand{\meanNch}{\mbox{$\langle N_{ch} \rangle$}}
\newcommand{\gev}{\mbox{$\mathrm{GeV}$}}
\newcommand{\gevc}{\mbox{${\mathrm{GeV/}}c$}}
\newcommand{\mevc}{\mbox{${\mathrm{MeV/}}c$}}
\newcommand{\TAA}{\mbox{$\mathrm{T}_{\mathrm{AA}}$}}
\newcommand{\siginel}{\mbox{$\sigma_{\mathrm{inel}}$}}

\begin{document}

\title{
Multiplicity distribution and spectra of negatively charged hadrons in
Au+Au collisions at $\sqrt{s_{_{\mathrm{NN}}}}$ = 130 GeV}

\author{
C.~Adler$^{11}$, Z.~Ahammed$^{23}$, C.~Allgower$^{12}$, J.~Amonett$^{14}$,
B.D.~Anderson$^{14}$, M.~Anderson$^5$, G.S.~Averichev$^{9}$, 
J.~Balewski$^{12}$, O.~Barannikova$^{9,23}$, L.S.~Barnby$^{14}$
% ...
\\
(STAR Collaboration)
}
\address{
% ...
$^5$University of California, Davis, California 95616\\
% ...
$^{9}$Laboratory for High Energy (JINR), Dubna, Russia\\
% ...
$^{11}$University of Frankfurt, Frankfurt, Germany\\
$^{12}$Indiana University, Bloomington, Indiana 47408\\
% ...
$^{14}$Kent State University, Kent, Ohio 44242\\
% ...
$^{23}$Purdue University, West Lafayette, Indiana 47907\\
% ...
}

\date{\today}
\maketitle


\begin{abstract}
    The minimum bias multiplicity distribution and the transverse
    momentum and pseudorapidity distributions for central collisions
    have been measured for negative hadrons (\hminus) in \AuAu\ 
    interactions at \sqrtsNN\,= 130 GeV. The multiplicity density at
    midrapidity for the 5\% most central interactions is
    $d\Nhminus/d\eta|_{\eta = 0} = 280 \pm 1(\textrm{stat})\pm
    20(\textrm{syst})$, an increase per participant of 38\% relative
    to \ppbar\ collisions at the same energy. The mean transverse
    momentum is $0.508\pm0.012$ GeV/c and is larger than in central \PbPb\ 
    collisions at lower energies. The scaling of the \hminus\ yield
    per participant is a strong function of \pt. The pseudorapidity
    distribution is almost constant within $|\eta|<1$.
\end{abstract}


\begin{multicols}{2}
    

The collision of high energy heavy ions is a promising laboratory for
the study of nuclear matter at high energy density and the possible
creation and decay of the Quark Gluon Plasma \cite{qm}. A new era in
the study of high energy nuclear collisions began in the year 2000
with the first operation of the Relativistic Heavy Ion Collider (RHIC)
at Brookhaven National Laboratory.

The multiplicity and inclusive single particle transverse momentum
(\pt) distributions of hadrons are important tools for understanding
the evolutionary path of the system created in the collision and help
to determine the characteristics of the early, hot and dense phase. In
this Letter we present the minimum-bias multiplicity distribution and
the pseudorapidity ($\eta$) and \pt\ distributions for the 5\% most
central collisions for negative hadrons (\hminus) in \AuAu\ 
interactions at a center-of-mass energy of \sqrtsNN\,= 130 GeV per
nucleon pair, measured by the STAR detector at RHIC. The results are
compared with reference data from \ppbar\ collisions at a similar
energy and collisions of heavy nuclei at a lower energy.


The main tracking detector for STAR is a large Time Projection Chamber
(TPC), which measures charged particles in the pseudorapidity range
$|\eta| < 1.8$ with complete azimuthal acceptance. It is placed inside
a uniform solenoidal magnetic field of strength 0.25 T.  The trigger
detectors are an array of scintillator slats (CTB) arranged in a
barrel surrounding the TPC, and two hadronic calorimeters
(ZDCs) at $\pm 18$ m from the detector center and at zero degrees
relative to the beam axis. The ZDCs intercept spectator neutrons from
the collision and provide a measure of the collision centrality.
Further details on the apparatus can be found in \cite{star}.

During the Summer 2000 run, RHIC delivered collisions between Au
nuclei at \sqrtsNN\,= 130 GeV. The data
presented here are from a minimum-bias sample, triggered by a
coincidence of signals above threshold in both ZDCs with the RHIC beam
crossing. The ZDC threshold was set to ensure efficient detection of
single spectator neutrons.  The efficiency of the ZDC coincidence
trigger for central events was measured using a high-threshold CTB
trigger.  The trigger efficiency was found to be above 99\% for the
entire range of multiplicities reported in this Letter.

The offline reconstruction found a primary vertex for each event by
propagating the measured tracks through the field towards the
beamline. The vertex resolution for high multiplicity events is
approximately 150 $\mu$m, both perpendicular and parallel to the beam
axis.  Events used in the analysis have a vertex within $\pm 95$ cm of
the center of the TPC along the beam axis.  The vertex finding
efficiency is 100\% for events with more than 50 primary tracks in the
TPC acceptance, decreasing to 60\% for those with fewer than 5 primary
tracks. 60,000 minimum-bias \AuAu\ events were used for this analysis.


Particle production was studied through the yield of primary negative
hadrons, comprising mostly \piminus\ with an admixture of
\kminus\ and \pbar.  The \hminus\ distribution includes the products
of strong and electromagnetic decays. Negatively charged hadrons were
studied in order to exclude effects due to participant nucleons.
Charged particle tracks reconstructed in the TPC were accepted for
this analysis if they fulfilled requirements on number of points on
the track and on pointing accuracy to the event vertex.  The measured
raw distributions were corrected for acceptance, track reconstruction
efficiency, and contamination due to interactions in material,
misidentified non-hadrons, and the products of weak decays. The
reconstruction efficiency was determined by embedding simulated tracks
into real events at the raw data level, reconstructing the full
events, and comparing the simulated input to the reconstructed output.
This technique requires a precise simulation of isolated single
tracks, achieved by a detailed simulation of the STAR apparatus based
on GEANT \cite{gstar} and a microscopic simulation of the TPC
response. The multiplicity of the embedded tracks was limited to 5\%
of the multiplicity of the real event in the same phase space as the
simulated data, thereby perturbing the real event at a level below the
statistical fluctuations within the event sample.

The acceptance is on average 95\% for tracks within the fiducial
volume having $\pt>300$ \mevc. The tracking efficiency ranges between
70\%--95\%, depending on \pt\ and the multiplicity of the event. For
tracks with $\pt>200$ \mevc\ the efficiency is above 85\%. Accepted
tracks for this analysis have $0.1 < \pt\,< 2$ \gevc\ and
$|\eta|<1.0$.
   
Instrumental backgrounds due to photon conversions and secondary
interactions with detector material were estimated using the detector
response simulations mentioned above, together with events generated
by the {\sc HIJING} model \cite{hijing}. The simulations were
calibrated using data in regions where background processes could be
directly identified.  The measured yield also contains contributions
from the products of weak decays, primarily \kzeros, that were
incorrectly reconstructed as primary tracks. The background fraction
of the raw signal is approximately 20\% at \pt\,= 100 \mevc,
decreasing with increasing \pt.  The average fraction of
background tracks in the uncorrected sample is 7\%. All corrections
were calculated as a function of the uncorrected event multiplicity.
The systematic uncertainty due to the corrections was estimated by
studying the variation in the final spectra due to both a large
variation in the track quality cuts with corresponding recalculation
of the correction factors, and a small variation in the track quality
cuts with no adjustment of the correction factors.

\begin{figure}[htb]
\begin{center}
    \includegraphics[width=8.6cm]{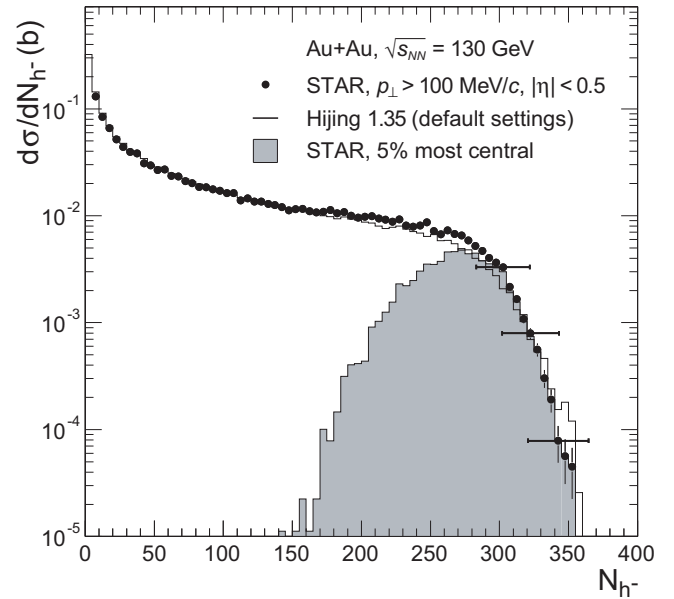}
       \caption{Normalized multiplicity distribution of \hminus\ with $\pt > 100\ \mevc$
           in \AuAu\ collisions at \sqrtsNN\, = 130 GeV.  The shaded
           area is 5\% most central collisions, selected by ZDC
           coincidence.  The solid curve is the prediction from the {\sc
               HIJING} model.}
        \label{fig:hminus}
    \end{center}
\end{figure}

Figure \ref{fig:hminus} shows the corrected, normalized multiplicity
distribution within $|\eta| < 0.5$ and $\pt > 100\ \mevc$ for minimum
bias \AuAu\ collisions. The data were normalized assuming a total
hadronic inelastic cross section of 7.2 barn for \AuAu\ collisions at
\sqrtsNN\,= 130 GeV, derived from Glauber model calculations
\cite{glauber}. The multiplicity bin below \Nhminus\,= 5 is not shown,
due to large systematic uncertainties in the vertex reconstruction
efficiency and large background contamination. Its relative
contribution to the total cross section was estimated to be 21\% by
normalizing the {\sc HIJING} multiplicity distribution to the measured
data in the region $5<\Nhminus<25$.  This procedure relies on the
assumption that very peripheral interactions are well described by the
superposition of a few nucleon-nucleon collisions in the geometry of a
nuclear collision, and can therefore be accurately modelled by {\sc
    HIJING}. The systematic error on the vertical scale is estimated
to be 10\% and is dominated by uncertainties in the total hadronic
cross section and the relative contribution of the first bin.  The
systematic error on the horizontal scale is 6\% for the entire range
of multiplicity and is depicted by horizontal error bars on a few data
points only.

The shape of the \hminus\ multiplicity distribution is dominated over
much of the \Nhminus\ range by the nucleus-nucleus collision geometry,
consistent with findings at lower energies.
However, the shape of the tail region at large \Nhminus\ is determined
by fluctuations and acceptance.  These overall features are also
observed in the HIJING calculation, shown as histogram in
Fig.~\ref{fig:hminus}.  The distribution for the 5\% most central
collisions (360 mb), defined using ZDC coincidence, is shown as the
shaded area in Fig.~\ref{fig:hminus}.

\begin{figure}[htb]
    \begin{center}
        \includegraphics[width=8.6cm]{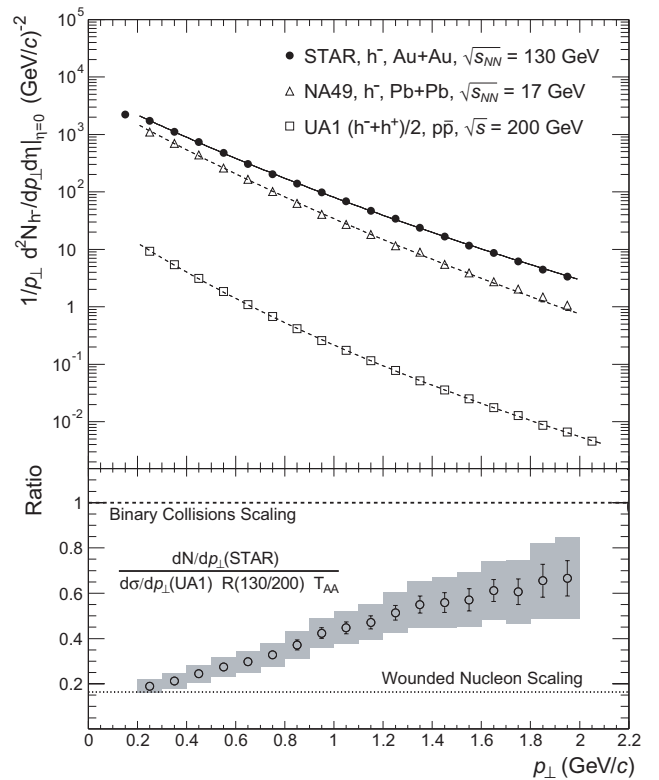}
        \caption{Upper panel: \hminus\ \pt-spectra for the 5\% most central \AuAu\ collisions
            at midrapidity ($|\eta| < 0.1$) for several systems. The
            curves are power-law fits to the data. Lower panel: ratio
            of STAR and scaled UA1 \pt-distributions (see text).}
        \label{fig:pt}
    \end{center}
\end{figure}

Figure \ref{fig:pt}, upper panel, shows the transverse momentum
distribution of negatively charged hadrons for central \AuAu\ 
collisions at midrapidity ($|\eta| < 0.1$) within $0.1 < \pt\, < 2\ 
\gevc$. Statistical errors are smaller than the symbols. The
correlated systematic error is estimated to be below 6\%. The data are
fit in the range $0.2 < \pt < 2$ \gevc\ by a QCD inspired power-law
function of the form $d^2\Nhminus/dp_\perp^2 d\eta = A\,
(1+\pt/p_0)^{-n}$ where $A$, $n$, and $p_0$ are free parameters.  The
upper panel of Fig.~\ref{fig:pt} also shows the \pt-distributions of
negatively charged hadrons for central \PbPb\ collisions at
\sqrtsNN\,= 17 GeV from NA49 \cite{na49} and for minimum-bias \ppbar\ 
collisions at \sqrts\,= 200 GeV from UA1 \cite{ua1}, fitted with the
same function.  The NA49 distribution, which was reported in units of
pion rapidity, was transformed to units of pseudorapidity.  The UA1
invariant cross section $E d^3 \sigma/d^3p$ reported in
Ref.~\cite{ua1} was scaled by $2 \pi / \siginel$, where $\siginel =
42$ mb \cite{ua5x}.  The power law fits all three datasets well.  The
mean \pt\ can be derived from the fit parameters as $\meanpt = 2 p_0 /
(n-3)$.  The fit to the STAR data gives $p_0 = 3.0 \pm 0.3\ \gevc$, $n
= 14.8 \pm 1.2$, and $\meanpt = 0.508 \pm 0.012\ \gevc$.  The strong
correlation of fit parameters $p_0$ and $n$ must be taken into account
when calculating the error on \meanpt.  The \meanpt\ from STAR is
larger than that from both central collisions of heavy nuclei at much
lower energy ($\meanpt_{\mathrm{NA49}} \simeq 0.429\ \gevc$
\cite{na49pt}) and nucleon-nucleon collisions at a comparable energy
($\meanpt_{\mathrm{UA1}} = 0.392 \pm 0.003\ \gevc$).

The PHENIX collaboration has reported that \meanEt/\meanNch\ for
central collisions of heavy nuclei is constant between SPS and RHIC
energies, with a value of $\sim 0.8$ GeV \cite{phenixEt}. Under the
assumption that the particle composition does not change significantly
between the SPS and RHIC, this finding is in apparent disagreement
with the above observation that \meanpt\ increases by 18\%. It should
be noted, however, that the systematic error on the SPS measurement of
\meanEt/\meanNch\ is 20\% \cite{wa98}, so that these results are
consistent within errors.

Figure \ref{fig:pt}, lower panel, shows the ratio of the STAR and UA1
\pt-distributions. Since the UA1 distribution is measured at \sqrts\,=
200 GeV, $d\sigma/d\pt$ is scaled by two factors for quantitative
comparison to the STAR data at 130 GeV: \textit{(i)} $R(130/200)$, the
\pt-dependent ratio of invariant cross sections for charged particle
production in \ppbar\ collisions at \sqrts\,= 130 and 200 GeV, and
\textit{(ii) } \TAA = $26\pm2$ mb$^{-1}$, the nuclear overlap integral
\cite{taa} for the 5\% most central \AuAu\ collisions.  $R$ varies from
0.92 at $\pt = 0.2\ \gevc$ to 0.70 at $\pt = 2.0\ \gevc$, and was
derived using scaling laws for \meanpt\ and $dN_{\mathrm{ch}}/d\eta$
as a function of \sqrts\ \cite{ua1,pdg} together with the
extrapolation to 130 GeV of power-law parameterizations at \sqrts\,=
200--900 GeV \cite{ua1}.  The shaded boxes show the total error of the
ratio, which is the linear sum of the errors of the measured data,
depicted by the error bars, and the systematic error due to
uncertainties in the scaling with \TAA\ and $R$.

There are two simple predictions for the scaled ratio. In lower energy
hadronic and nuclear collisions, the total pion yield due to soft (low
\pt) processes scales as the number of participants (``wounded''
nucleons) in the collision (e.g.~\cite{na49,bialas}).  The scaled
ratio in this case is 0.164, assuming 172 participant pairs
\cite{dima} and a mean number of binary collisions of 1050
($=\siginel\, \TAA$, where for $\sqrts = 130\,\gev\ \siginel \simeq
40.5$ mb \cite{ua5x}) for the 5\% most central \AuAu\ events.  In
contrast, if hadron production is due to hard (high \pt) processes and
there are no nuclear-specific effects (see below), the hadron yield
will scale as the number of binary nucleon-nucleon interactions in the
nuclear collision and the value of the ratio is unity.  There are
important nuclear effects which may alter the scaling as a function of
\pt\ from these simple predictions, including initial state multiple
scattering \cite{cronin}, shadowing \cite{shadowing}, jet quenching
\cite{jetq}, and radial flow \cite{rflow}.  The scaled ratio exhibits
a strong \pt\ dependence, rising monotonically with increasing \pt\ 
from Wounded Nucleon scaling at low \pt\ but not reaching Binary
Collision scaling at the highest \pt\ reported. This is consistent
with the presence of radial flow, as well as the onset of hard
scattering contributions and initial state multiple scattering with
rising \pt.

\begin{figure}[p]
    \begin{center}
        \includegraphics[width=8.6cm]{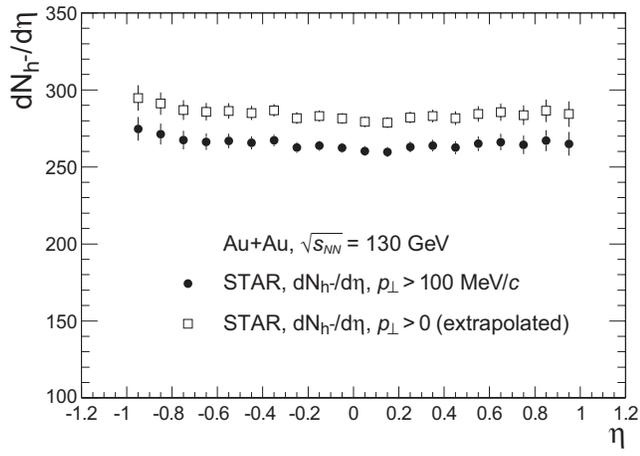}
        \caption{\hminus\ pseudorapidity distribution
            from 5\% most central \AuAu\ collisions for $\pt\,> 100$
            \mevc\ (filled circles) and all \pt\ (open squares).}
        \label{fig:dndeta}
    \end{center}
\end{figure}

Figure \ref{fig:dndeta} shows the normalized pseudorapidity
distribution of \hminus\ for the 5\% most central collisions within
$|\eta| < 1.0$, both for $\pt\ > 100\ \mevc$ and for all \pt. The
latter was obtained by fitting a power-law function in the range $0.1
< \pt < 2\ \gevc$ and extrapolating to $\pt = 0$ in order to estimate
the content of the first bin.  The error bars indicate the
uncorrelated systematic errors. The statistical errors are negligible.
The correlated systematic error applied to the overall normalization
is estimated to be below 6\% for $\pt\ > 100$ \mevc\ and 7\% for all
\pt.

The $\eta$ distribution is almost constant within $|\eta|<1$,
exhibiting a small rise at larger $\eta$. This shape is expected from
a boost invariant source (i.e., constant in rapidity), taking into
account the transformation from $y$ to $\eta$.
It should be noted that in \PbPb\  collisions at \sqrtsNN\,= 17 GeV the
pseudo-rapidity distribution of charged hadrons \cite{na45} and the
rapidity distribution of negative hadrons (assuming the pion mass)
\cite{na49} were found to peak at midrapidity, suggesting a significant
change in the longitudinal phase space distribution between the SPS and
RHIC. Measurement of the rapidity distribution of identified particles
is needed to establish the boost invariance at RHIC.

The \hminus\ density at midrapidity for $\pt > 100\, \mevc$ is
$dN/d\eta|_{\eta = 0} = 261 \pm 1(\mbox{stat})\pm 17(\mbox{syst})$.
Extrapolation to $\pt$ = 0 yields $dN/d\eta|_{\eta = 0} = 280 \pm
1(\mbox{stat})\pm 20(\mbox{syst})$. Assuming an average of 172
participant pairs per central \AuAu\ collision, this corresponds to
$1.63 \pm 0.12$\,\hminus\ per participant nucleon pair per unit
pseudorapidity, a 38\% increase over the yield in \ppbar\ collisions
extrapolated to the same energy \cite{ua5} (we neglect isospin
correction factors of order 1--3\%) and a 52\% increase over \PbPb\ 
collisions at \sqrtsNN\,= 17 GeV \cite{na49}.

For the total charged multiplicity density in \AuAu\ interactions at
\sqrtsNN\,= 130 GeV, the PHOBOS collaboration has reported
$dN_{ch}/d\eta|_{|\eta|<1} = 555 \pm 12(\mathrm{stat}) \pm
35(\mathrm{syst})$ (6\% most central collisions) \cite{phobos}, while
the PHENIX collaboration has reported $dN_{ch}/d\eta|_{\eta=0} = 622
\pm 1(\mathrm{stat}) \pm 41(\mathrm{syst})$ (5\% most central)
\cite{phenix}.  To compare to these results, positive charged
particles were analysed in the framework described above. For the 5\%
most central collisions, STAR measures a total charged multiplicity
density of $dN_{ch}/d\eta|_{\eta=0} = 567 \pm 1(\mathrm{stat}) \pm
38(\mathrm{syst})$.


In conclusion, we find that particle production per participant in
central \AuAu\ collisions at \sqrtsNN\,=130 GeV increases by 38\%
relative to \ppbar\ and 52\% compared to nuclear collisions at
\sqrtsNN\,= 17 GeV. The \pt\ distribution is harder than that of the
\ppbar\ reference system for the \pt\ region up to 2 \gevc.  Scaling of
produced particle yield with number of participants shows a strong
dependence on \pt, with Wounded Nucleon scaling achieved only at the
lowest measured \pt. The \hminus\ pseudorapidity distribution is
almost constant within $|\eta|<1$.


Acknowledgments: We wish to thank the RHIC Operations Group at
Brookhaven National Laboratory for their tremendous support. This work
was supported by the Division of Nuclear Physics and the Division of
High Energy Physics of the Office of Science of the U.S.Department of
Energy, the United States National Science Foundation, the
Bundesministerium fuer Bildung und Forschung of Germany, the Institut
National de la Physique Nucleaire et de la Physique des Particules of
France, the United Kingdom Engineering and Physical Sciences Research
Council, Fundacao de Amparo a Pesquisa do Estado de Sao Paulo, Brazil,
and the Russian Ministry of Science and Technology.


\end{multicols}
\end{document}